# Machine learning for neural decoding

Joshua I. Glaser[1,2,6,8,9*], Ari S. Benjamin[6], Raeed H. Chowdhury[3,4], Matthew G. Perich[3,4], Lee E. Miller[2-4], and Konrad P. Kording[2-7]

1. Interdepartmental Neuroscience Program, Northwestern University, Chicago, IL, USA
2. Department of Physical Medicine and Rehabilitation, Northwestern University and Shirley Ryan Ability Lab, Chicago, IL, USA
3. Department of Physiology, Northwestern University, Chicago, IL, USA
4. Department of Biomedical Engineering, Northwestern University, Chicago, IL, USA
5. Department of Applied Mathematics, Northwestern University, Chicago, IL, USA
6. Department of Bioengineering, University of Pennsylvania, Philadelphia, IL, USA
7. Department of Neuroscience, University of Pennsylvania, Philadelphia, IL, USA
8. Department of Statistics, Columbia University, New York, NY, USA
9. Zuckerman Mind Brain Behavior Institute, Columbia University, New York, NY, USA

* Contact: joshglaser88@gmail.com

## Abstract
Despite rapid advances in machine learning tools, the majority of neural decoding approaches still use traditional methods. Modern machine learning tools, which are versatile and easy to use, have the potential to significantly improve decoding performance. This tutorial describes how to effectively apply these algorithms for typical decoding problems. We provide descriptions, best practices, and code for applying common machine learning methods, including neural networks and gradient boosting. We also provide detailed comparisons of the performance of various methods at the task of decoding spiking activity in motor cortex, somatosensory cortex, and hippocampus. Modern methods, particularly neural networks and ensembles, significantly outperform traditional approaches, such as Wiener and Kalman filters. Improving the performance of neural decoding algorithms allows neuroscientists to better understand the information contained in a neural population and can help advance engineering applications such as brain machine interfaces.

## Introduction
Neural decoding uses activity recorded from the brain to make predictions about variables in the outside world. For example, researchers predict movements based on activity in motor cortex (Serruya et al., 2002; Ethier et al., 2012), decisions based on activity in prefrontal and parietal cortices (Baeg et al., 2003; Ibos and Freedman, 2017), and spatial locations based on activity in the hippocampus (Zhang et al., 1998; Davidson et al., 2009). These decoding predictions can be used to control devices (e.g., a robotic limb), or to better understand how areas of the brain relate to the outside world. Decoding is a central tool in neural engineering and for neural data analysis.

In essence, neural decoding is a regression (or classification) problem relating neural signals to particular variables. When framing the problem in this way, it is apparent that there is a wide range of methods that one could apply. However, despite the recent advances in machine learning (ML) techniques for regression, it is still common to decode activity with traditional methods such as linear regression. Using modern ML tools for neural decoding has the potential to boost performance significantly and might allow deeper insights into neural function.

This tutorial is designed to help readers start applying standard ML methods for decoding. We describe when one should (or should not) use ML for decoding, how to choose an ML method, and best practices such as cross-validation and hyperparameter optimization. We provide companion code that makes it possible to implement a variety of decoding methods quickly. Using this same code, we demonstrate here that ML methods outperform traditional decoding methods. In example datasets of recordings from monkey motor cortex, monkey somatosensory cortex, and rat hippocampus, modern ML methods showed the highest accuracy decoding of available methods. Using our code and this tutorial, readers can achieve these performance improvements on their own data.

## Materials and Methods

In this section, we provide 1) general background about decoding, including when to use machine learning (ML) for decoding and considerations for choosing a decoding method; 2) the practical implementation details of using ML for decoding, including data formatting, proper model testing, and hyperparameter optimization; and 3) methods that we use to compare ML techniques in the *Results* section, including descriptions of the specific ML techniques we compare and datasets we use.

## When to use machine learning for decoding

Machine learning is most helpful when the central research aim is to obtain greater predictive performance. This is in part because of the general success of machine learning for nonlinear problems (He et al., 2015; LeCun et al., 2015; Silver et al., 2016). We will demonstrate this later in this tutorial in *Results*. There are multiple separate research aims for decoding that benefit from improved predictive performance, including engineering applications and, if used carefully, understanding neural activity (Glaser et al., 2019).

### *Engineering applications*

Decoding is often used in engineering contexts, such as for brain machine interfaces (BMIs), where signals from motor cortex are used to control computer cursors (Serruya et al., 2002), robotic arms (Collinger et al., 2013), and muscles (Ethier et al., 2012). When the primary aim of these engineering applications is to improve predictive accuracy, ML should generally be beneficial.

### *Understanding what information is contained in neural activity*

Decoding is also an important tool for understanding how neural signals relate to the outside world. It can be used to determine how much information neural activity contains about an external variable (e.g., sensation or movement) (Hung et al., 2005; Raposo et al., 2014; Rich and Wallis, 2016), and how this information differs across brain areas (Quiroga et al., 2006; Hernández et al., 2010; van der Meer et al., 2010), experimental conditions (Dekleva et al., 2016; Glaser et al., 2018), and disease states (Weygandt et al., 2012). When the goal is to determine how much information a neural population has about an external variable, regardless of the form of that information, then using ML will generally be beneficial. However, when the goal is to determine how a neural population processes that information, or to obtain an interpretable description of how that information is represented, one should exercise care with ML, as we describe in the next section.

### *Benchmarking for simpler decoding models*

Decoders can also be used to understand the form of the mapping between neural activity and variables in the outside world (Pagan et al., 2016; Naufel et al., 2018). That is, if researchers aim to test if the mapping from neural activity to behavior/stimuli (the "neural code") has a certain structure, they can develop a "hypothesis-driven decoder" with a specific form. If that decoder can predict task variables with some arbitrary accuracy level, this is sometimes held as evidence that information within neural activity indeed has the hypothesized structure. However, it is important to know how well a hypothesis-driven decoder performs relative to what is possible. This is where modern ML methods can be of use. If a method designed to test a hypothesis decodes activity much worse than ML methods, then a researcher knows that their hypothesis likely misses key aspects of the neural code. Hypothesis-driven decoders should thus always be compared against a good-faith effort to maximize performance accuracy with a good machine learning approach.

## Caution in interpreting machine learning models of decoding

### *Understanding how information in neural activity relates to external variables*

It is tempting to investigate how an ML decoder, once fit to neural data, transforms neural activity to external variables. This may be especially tempting if the ML model resembles neural function, such as in a neural network decoder. Still, high predictive performance is not evidence that transformations occurring within the ML decoder are the same, or even similar to, those in the brain. In general, and unlike hypothesis-driven decoders, the mathematical transformations of most ML decoders are hard to interpret and themselves are not meant to represent any specific biological variable. Some recent efforts have started to investigate how ML models might be interpreted once fit to data (Ribeiro et al., 2016; Olah et al., 2018; Cranmer et al., 2020). However, users should be cautious that this is an active research area and that, in general, ML methods are not designed for mechanistic interpretation.

### *Understanding what information is contained in neural activity*

It is important to be careful with the scientific interpretation of decoding results, both for ML and other models (Weichwald et al., 2015). Decoding can tell us how much information a neural population has about a variable $X$. However, high decoding accuracy does not mean that a brain area is directly involved in processing $X$, or that $X$ is the purpose of the brain area (Wu et al., 2020). For example, with a powerful decoder, it could be possible to accurately classify images based on recordings from the retina, since the retina has information about all visual space. However, this does not mean that the primary purpose of the retina is image classification. Moreover, even if the neural signal temporally precedes the external variable, it does not necessarily mean that it is causally involved (Weichwald et al., 2015). For example, movement-related information could reach somatosensory cortex prior to movement due to an efference copy from motor cortex, rather than somatosensory cortex being responsible for movement generation. Researchers should constrain interpretations to address the information in neural populations about variables of interest, but not use this as evidence for areas' roles or purpose.

On a more technical level, there are some decoders that will not directly tell us what information is contained in the neural population. As described more in the following section, some neural decoders incorporate prior information, e.g. incorporating the overall probability of being in a given location when decoding from hippocampal place cells (Zhang et al., 1998). If a decoder uses prior information about the decoded variable, then the final decoded variables will not only reflect the information contained in the neural population, but will also reflect the prior information - the two will be entangled (Kriegeskorte and Douglas, 2019).

## What decoder should I use to improve predictive performance?

Depending on the recording method, location, and variables of interest, different decoding methods may be most effective. Neural networks, gradient boosted trees, support vector machines, and linear methods are among the dozens of potential candidates. Each makes different assumptions about how inputs relate to outputs. We will describe a number of specific methods suitable for neural decoding later in this tutorial. Ultimately, we recommend testing multiple methods, perhaps starting with the methods we have found to work best for our demonstration datasets. Still, it is important to have a general understanding of differences between methods.

***Different methods make different implicit assumptions about the data***

There is no such thing as a method that makes no assumptions. This idea derives from a key theorem in the ML literature called the "No Free Lunch" theorem, which essentially states that no algorithm will outperform all others on every problem (Wolpert and Macready, 1997). The fact that some algorithms perform better than others in practice means that their assumptions about the data are better. The knowledge that all methods make assumptions to varying degrees can help one be intentional about choosing a decoder.

The assumptions of some methods are very clear. Regularized (ridge, or $L_2$) linear regression, for example, makes three: the change in the outputs is assumed to be proportionate to the change in the inputs, any additional noise on the outputs is assumed to be Gaussian noise (implied by the mean-squared error), and the regression coefficients are assumed to pull from a Gaussian distribution (Hoerl and Kennard, 1970; Bishop, 2006). For more complicated methods like neural networks, the assumptions are more complicated (and still under debate, e.g. (Arora et al., 2019)) but still exist.

One crucial assumption that is built into decoders is the form of the input/output relation. Some methods, e.g. linear regression or a Kalman filter, assume this mapping has a fixed linear form. Others, e.g. neural networks, allow this mapping to have a flexible nonlinear form. In the scenario that the mapping is truly linear, making this linear assumption will be beneficial; it will be possible to accurately learn the decoder when there is less data or more noise (Hastie et al., 2009). However, if the neural activity relates nonlinearly to the task variable, then it can be beneficial to have a decoder that is relatively agnostic about the overall input/output relationship, in the sense that it can express many different types of relationships.

A challenge of choosing a model that is highly expressive or complex is the phenomenon of "overfitting", in which the model learns components of the noise that are unique to the data that the model is trained on, but which do not generalize to any independent test data. To combat overfitting, one can choose a simpler algorithm that is less likely to learn to model this noise. Alternatively, one can apply regularization, which essentially penalizes the complexity of a model. This effectively reduces the expressivity of a model, while still allowing it to have many free parameters (Friedman et al., 2001). Regularization is an important way to include the assumption that the input/output relationship is not arbitrarily complex.

Additionally, for some neural decoding problems, it is possible to incorporate assumptions about the output being predicted. For example, let us say we are estimating the kinematics of a movement from neural activity. The Kalman filter, which is frequently used for movement decoding (Wu et al., 2003; Wu and Hatsopoulos, 2008; Gilja et al., 2012), uses the additional information that movement kinematics transition over time in a smooth and orderly manner. There are also Bayesian decoders that can incorporate prior beliefs about the decoded variables. For example, the Naive Bayes decoder, which is frequently used for position decoding from hippocampal activity (Zhang et al., 1998; Barbieri et al., 2005; Kloosterman et al., 2013), can take into account prior information about the probability distribution of positions. Decoders with task knowledge built in are constrained in terms of their solutions, which can be helpful, provided this knowledge is correct.

### *Maximum likelihood estimates vs. posterior distributions*

Different classes of methods also provide different types of estimates. Typically, machine learning algorithms provide maximum likelihood estimates of the decoded variable. That is, there is a single point estimate for the value of the decoded variable, which is the estimate that is most likely to be true. Unlike the typical use of machine learning algorithms, Bayesian decoding provides a probability distribution over all possibilities for the decoded outputs (the "posterior" distribution), thus also providing information about the uncertainty of the estimate (Bishop, 2006). The maximum likelihood estimate is the peak of the posterior distribution when the prior distribution is uniform.

### *Ensemble methods*

It is important to note that the user does not need to choose a single model. One can usually improve on the performance of any single model by combining multiple models, creating what is called an "ensemble". Ensembles can be simple averages of the outputs of many models, or they can be more complex combinations. Weighted averages are common, which one can imagine as a second-tier linear model that takes the outputs of the first-tier models as inputs (sometimes referred to as a super learner). In principle, one can use any method as the second-tier model. Third- and fourth-tier models are uncommon but imaginable. Many successful ML methods are, at their core, ensembles (Liaw and Wiener, 2002). In ML competition sites like Kaggle (Kaggle, 2020), most winning solutions are complicated ensembles.

### *Under which conditions will ML techniques improve decoding performance?*

Whether ML will outperform other methods depends on many factors. The form of the underlying neural code, the length of the recording session, and the level of noise will all affect the predictive accuracy of different methods to different degrees. There is no way to know ahead of time which method will perform the best, and we recommend creating a pipeline to quickly test and compare many ML and simpler regression methods. For typical decoding situations, however, we expect certain modern ML methods to perform better than simpler methods. In the demonstration that follows in *Results*, we show that this is true across three datasets and a wide range of variables like training data length, number of neurons, bin size, and hyperparameters.

## A practical guide ("the nitty gritty") for using machine learning for decoding

In any decoding problem, one has neural activity from multiple sources that is recorded for a period of time. While we focus here on spiking neurons, the same methods could be used with other forms of neural data, such as the BOLD signal in fMRI, or the power in particular frequency bands of local field potential (LFP) or electroencephalography (EEG) signals. When we decode from the neural data, whatever the source, we would like to predict the values of recorded outputs (Fig. 1a).

### *Data formatting/preprocessing for decoding*

#### *Preparing for regression vs. classification*

Depending on the task, the desired output can be variables that are continuous (e.g., velocity or position), or discrete (e.g., choices). In the first case the decoder will perform regression, while in the second case it will perform classification. In our Python package, decoder classes are labeled to reflect this division.

In the data processing step, take note whether a prediction is desired continuously in time, or only at the end of each trial. In this tutorial, we focus on situations in which a prediction is desired continuously in time. However, many classification situations require only one prediction per trial (e.g., when making a single choice per trial). If this is the case, the data must be prepared such that many timepoints in a single trial are mapped to a single output.

*Time binning divides continuous data into discrete chunks*

A number of important decisions arise from the basic problem that time is continuously recorded, but decoding methods generally require discrete data for their inputs (and, in the continual-prediction situation, their outputs). A typical solution is to divide both the inputs and outputs into discrete time bins. These bins usually contain the average input or output over a small chunk of time, but could also contain the minimum, maximum, or the regular sampling of any interpolation method fit to the data.

When predictions are desired continuously in time, one needs to decide upon the temporal resolution, $R$, for decoding. That is, do we want to make a prediction every 50ms, 100ms, etc? We need to put the input and output into bins of length $R$ (Fig. 1a). It is common (although not necessary) to use the same bin size for the neural data and output data, and we do so here. Thus, if $T$ is the length of the recording, we will have approximately $T/R$ total data points of neural activity and outputs.

Next, we need to choose the time period of neural activity used to predict a given output. In the simplest case, the activity from all neurons in a given time bin would be used to predict the output in that same time bin. However, it is often the case that we want the neural data to precede the output (e.g., in the case of making movements) or follow the decoder output (e.g., in the case of inferring the cause of a sensation). Plus, we often want to use neural data from more than one bin (e.g., using 500 ms of preceding neural data to predict a movement in the current 50 ms bin). In the following, we use the nomenclature that $B$ time bins of neural activity are being used to predict a given output. For example, if we use three bins preceding the output and one concurrent bin, then $B$=4 (Fig. 1a). Note that when multiple bins of neural data are used to predict an output ($B$>1), then overlapping neural data will be used to predict different output times (Fig. 1a), making them statistically dependent.

When multiple bins of neural data are used to predict an output, then we will need to exclude some output bins. For instance, if we are using one bin of neural data preceding the output, then we cannot predict the first output bin, and if we are using one bin of neural data following the output, then we cannot predict the final output bin (Fig. 1a). Thus, we will be predicting $K$ total output bins, where $K$ is less than the total number of bins ($T/R$). To summarize, our decoders will be predicting each of these $K$ outputs using $B$ surrounding bins of activity from $N$ neurons.

*The format of the input data depends on the form of the decoder*

**Non-recurrent decoders:** For most standard regression methods, the decoder has no persistent internal state or memory. In this case $N$ x $B$ features (the firing rates of each neuron in each relevant time bin) are used to predict each output (Fig. 1b). The input matrix of covariates, $\boldsymbol{X}$, has $N$ x $B$ columns (one for each feature) and $K$ rows (corresponding to each output being predicted). If there is a single output that is being predicted, it can be put in a vector, $\boldsymbol{Y}$, of length $K$. Note that for many decoders, if there are multiple outputs, each is independently decoded. If multiple outputs are being simultaneously predicted, which can occur with neural network decoders, the outputs can be put in a matrix $\boldsymbol{Y}$, that has $K$ rows and $d$ columns, where $d$ is the number of outputs being predicted. Since this is the format of a standard regression problem, many regression methods can easily be substituted for one another once the data has been prepared in this way.

**Recurrent neural network decoders**: When using recurrent neural networks (RNNs) for decoding, we need to put the inputs in a different format. Recurrent neural networks explicitly model temporal transitions across time with a persistent internal state, called the "hidden state" (Fig. 1c). At each time of the $B$ time bins, the hidden state is adjusted as a function of both the $N$ features (the firing rates of all neurons in that time bin) and the hidden state at the previous time bin (Fig. 1c). Note that an alternative way to view this model is that the hidden state feeds back on itself (across time points). After transitioning through all $B$ bins, the hidden state in this final bin is used to predict the output. In this way an RNN decoder can integrate the effect of neural inputs over an extended period of time. For use in this type of decoder, the input can be formatted as a 3-dimensional tensor of size $K$ x $N$ x $B$ (Fig. 1c). That is, for each row (corresponding to one of the $K$ output bins to be predicted), there will be $N$ features (2nd tensor dimension) over $B$ bins (3rd tensor dimension) used for prediction. Within this format, different types of RNNs, including those more sophisticated than the standard RNN shown in Fig. 1c, can be easily switched for one another.

**a** Decoding Schematic

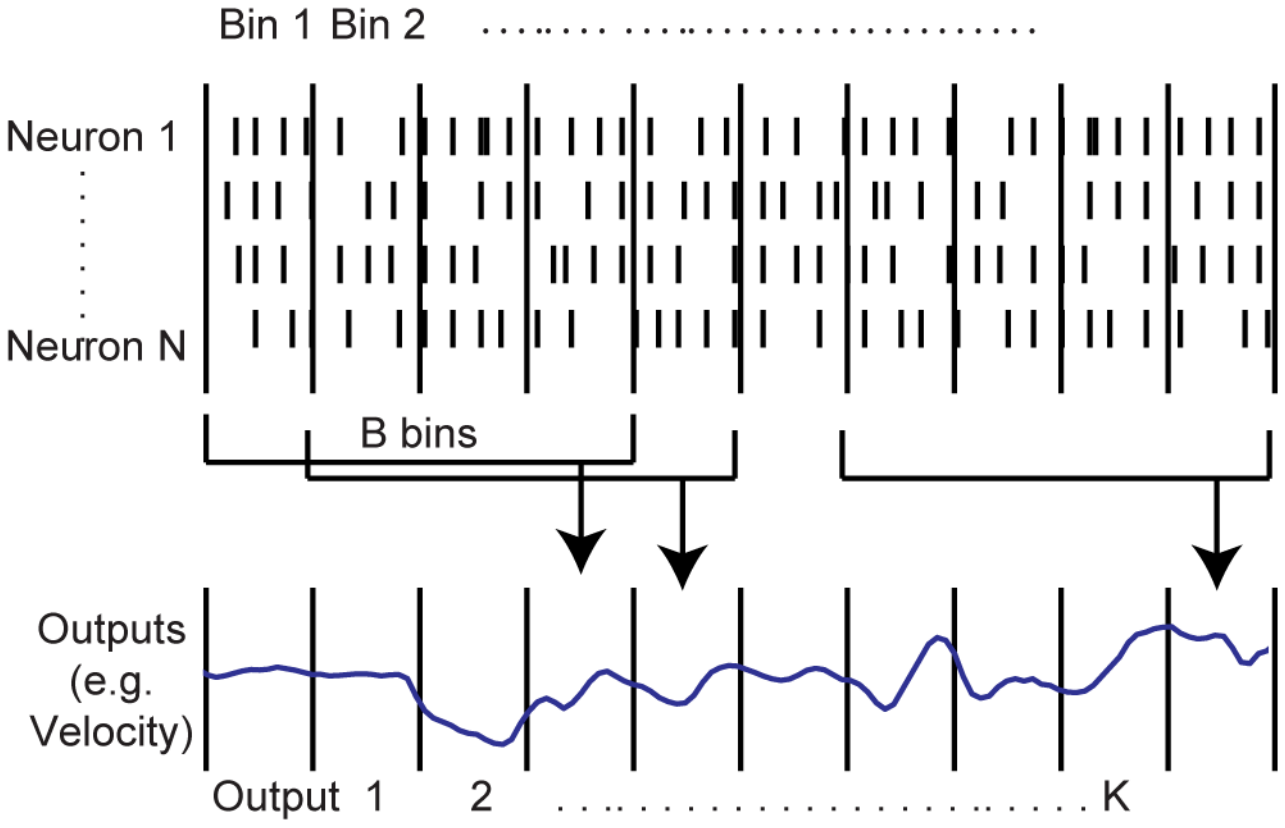

**b** Non-recurrent Decoders

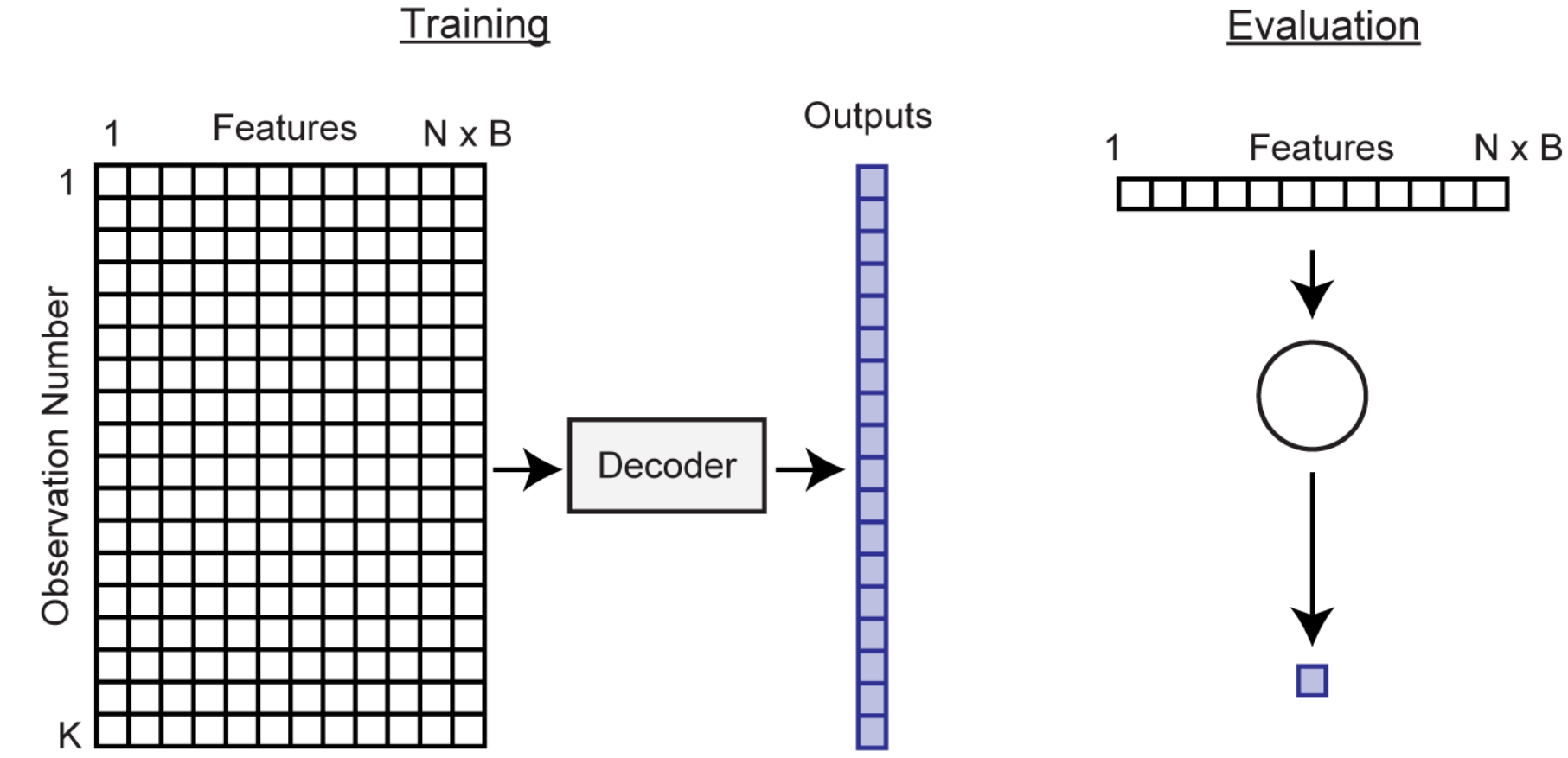

**c** Recurrent Decoders

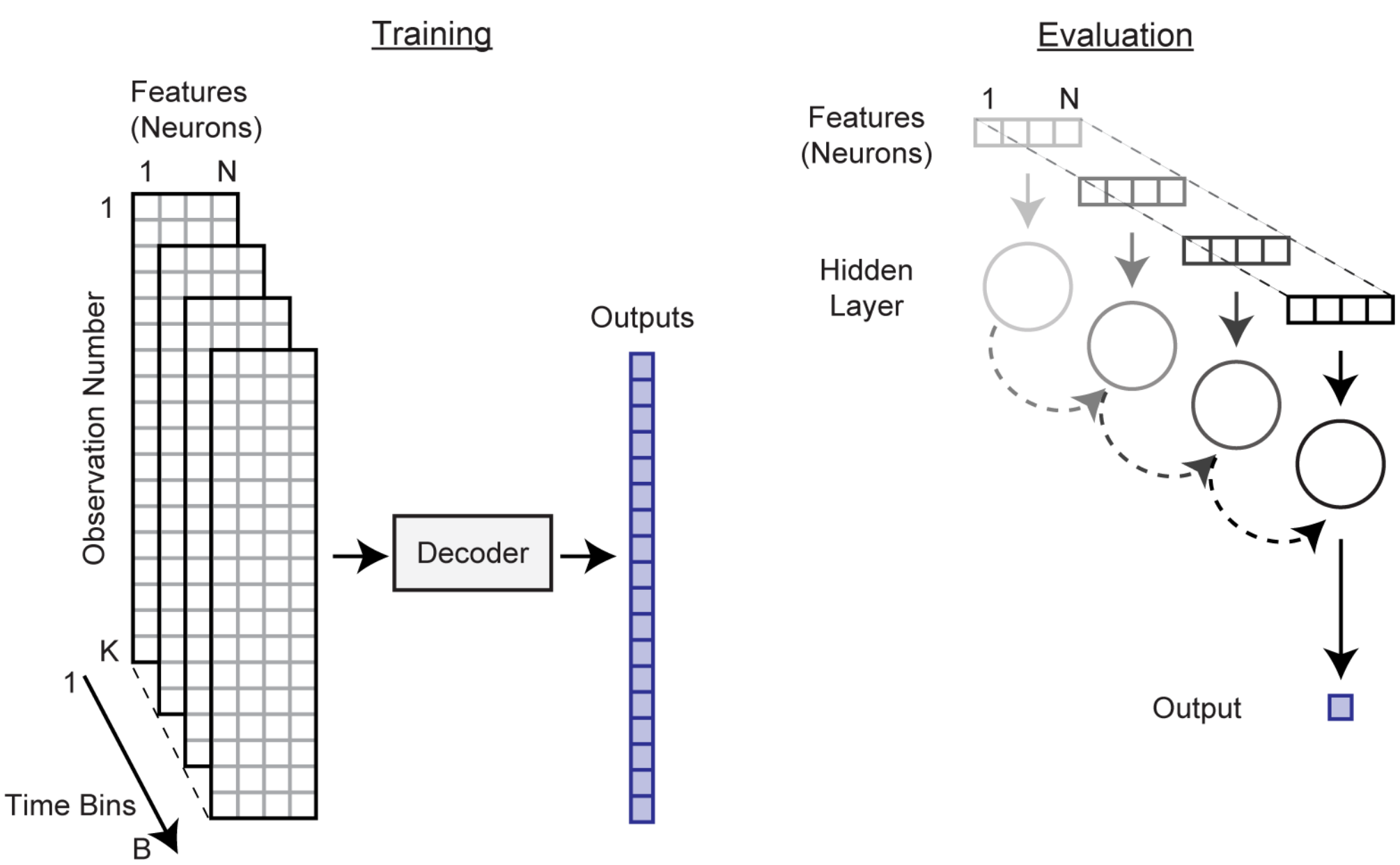

**Figure 1: Decoding Schematic**

**a)** To decode (predict) the output in a given time bin, we used the firing rates of all $N$ neurons in $B$ time bins. In this schematic, $N$=4 and $B$=4 (three bins preceding the output and one concurrent bin). As an example, preceding bins of neural activity could be useful for predicting upcoming movements and following bins of neural activity could be useful for predicting preceding sensory information. Here, we show a single output being predicted. **b)** For non-recurrent decoders (Wiener Filter, Wiener Cascade, Support Vector Regression, XGBoost, and Feedforward Neural Network in our subsequent demonstration), this is a standard machine learning regression problem where $N$ x $B$ features (the firing rates of each neuron in each relevant time bin) are used to predict the output. **c)** To predict outputs with recurrent decoders (simple recurrent neural network, GRUs, LSTMs in our subsequent demonstration) we used $N$ features, with temporal connections across $B$ bins. A schematic of a recurrent neural network predicting a single output is on the right. Note that an alternative way to view this model is that the hidden state feeds back on itself (across time points).

### *Applying machine learning*

The typical process of applying ML to data involves testing several methods and seeing which works best. Some methods may be expected to work better or worse depending on the structure of the data, and in the next part of this tutorial we provide a demonstration of which methods work best for typical neural spiking datasets. That said, applying ML is unavoidably an iterative process, and for this reason we begin with the proper way to choose among many methods. It is particularly important to consider how to avoid overfitting our data during this iterative process.

#### Compare method performance using cross-validation

Given two (or more) methods, the ML practitioner must have a principled way to decide which method is best. It is crucial to test the decoder performance on a separate, held-out dataset that the decoder did not see during training (Fig. 2a). This is because a decoder typically will *overfit* to its training data. That is, it will learn to predict outputs using idiosyncratic information about each datapoint in the training set, such as noise, rather than the aspects that are general to the data. An overfit algorithm will describe the training data well but will not be able to provide good predictions on new datasets. For this reason, the proper metric to choose between methods is the performance on held-out data, meaning that the available data should be split into separate "training" and "testing" datasets. In practice, this will reduce the amount of training data available. In order to efficiently use all of the data, it is common to perform cross-validation (Fig. 2b). In 10-fold cross-validation, for example, the dataset is split into 10 sets. The decoder is trained on 9 of the sets, and performance is tested on the final set. This is done 10 times in a rotating fashion, so that each set is tested once. The performance on all test sets is generally averaged together to determine the overall performance.

When publishing about the performance of a method, it is important to show the performance on held-out data that was additionally not used to select between methods (Fig. 2c). Random noise may lead some methods to perform better than others, and it is possible to fool oneself and others by testing a great number of methods and choosing the best. In fact, one can entirely overfit to a testing dataset simply by using the test performance to select the best method, even if no method is trained on that data explicitly. Practitioners often make the distinction between the "training" data, the "validation" data (which is used to select between methods) and the "testing" data (which is used to evaluate the true performance). This three-way split of data complicates cross-validation somewhat (Fig. 2b). It is possible to first create a separate testing set, then choose among methods on the remaining data using cross-validation. Alternatively, for maximum data efficiency, one can iteratively rotate which split is the testing set, and thus perform a two-tier nested cross-validation.

### *Hyperparameter optimization*

When fitting any single method, one usually has to additionally choose a set of *hyperparameters*. These are parameters that relate to the design of the decoder itself and should not be confused with the 'normal' parameters that are fit during optimization, e.g., the weights that are fit in linear regression. For example, neural networks can be designed to have any number of hidden units. Thus, the user needs to set the number of hidden units (the hyperparameter) before training the decoder. Often decoders have multiple hyperparameters, and different hyperparameter values can sometimes lead to greatly different performance. Thus, it is important to choose a decoder's hyperparameters carefully.

When using a decoder that has hyperparameters, one should take the following steps. First, always split the data into three separate sets (training set, testing set, and validation set), perhaps using nested cross-validation (Fig. 2b). Next, iterate through a large number of hyperparameter settings and choose the best based on validation set performance. Simple methods for searching through hyperparameters are a grid search (i.e., sampling values evenly) and random search (Bergstra and Bengio, 2012). There are also more efficient methods (e.g., (Snoek et al., 2012; Bergstra et al., 2013)) that can intelligently search through hyperparameters based on the performance of previously tested hyperparameters. The best performing hyperparameter and method combination (on the validation set) will be the final method, unless one is combining multiple methods into an ensemble.

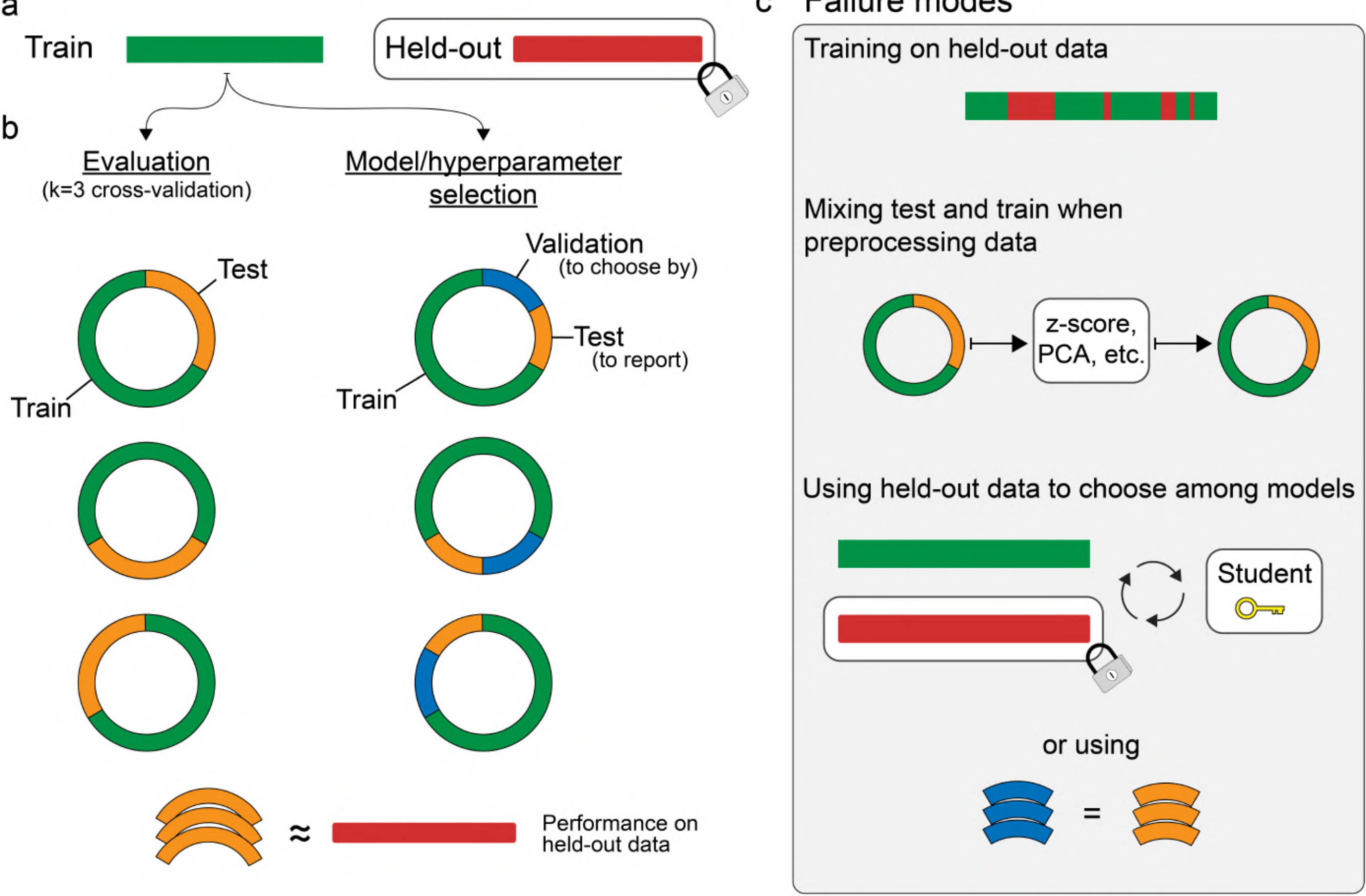

**Figure 2. Schematic of cross-validation**

**a)** After training a decoder on some data (green), we would like to know how well it performs on held-out data that we do not have access to (red). **b) Left:** By splitting the data we have into test (orange) and train (green) segments, we can approximate the performance on held-out data. In k-fold cross-validation, we re-train the decoder k times, and each time rotate which parts of the data are the test or train data. The average test set performance approximates the performance on held-out data. **Right:** If we want to select among many models, we cannot maximize the performance on the same data we will report as a score. (This is very similar to "p-hacking" statistical significance.) Instead we maximize performance on "validation" data (blue), and again rotate through the available data. **c)** All failure modes are ways in which a researcher lets information from the test set "leak" into the training algorithm. This happens if you explicitly train on the test data (top), or use any statistics of the test data to modify the train data before fitting (middle), or select your models or hyperparameters based on the performance on the test data (bottom).

## Methods for our decoding comparisons and demonstrations

In this section, we demonstrate the usage of our code package and show which of its methods work better than traditional methods for several neural datasets.

### *Code accessibility*

We have prepared a Python package that implements the machine learning pipeline for decoding problems. It is available at https://github.com/KordingLab/Neural_Decoding, and it includes code to correctly format the neural and output data for decoding, to implement many different decoders for both regression and classification, and to optimize their hyperparameters.

### *Specific decoders:*

The following decoders are available in our package, and we demonstrate their performance on example datasets below. We included both historical linear techniques (e.g., the Wiener filter) and modern ML techniques (e.g., neural networks and ensembles of techniques). Additional details of the methods, including equations and hyperparameter information, can be found in a table in *Extended Data* (Figure 4-3).

**Wiener Filter:** The Wiener filter uses multiple linear regression to predict the output from multiple time bins of all neurons' spikes (Carmena et al., 2003). That is, the output is assumed to be a linear mapping of the number of spikes in the relevant time bins from every neuron (Fig. 1a,b).
**Wiener Cascade:** The Wiener cascade (also known as a linear-nonlinear model) fits a linear regression (the Wiener filter) followed by a fitted static nonlinearity (e.g., (Pohlmeyer et al., 2007)). This allows for a nonlinear input-output relationship and assumes that this nonlinearity is purely a function of the linear output. Thus, there is no nonlinear mixing of input features when making the prediction. The default nonlinear component is a polynomial with degree that can be determined on a validation set.
**Support Vector Regression:** In support vector regression (SVR) (Smola and Schölkopf, 2004; Chang and Lin, 2011), the inputs are projected into a higher-dimensional space using a nonlinear kernel, and then linearly mapped from this space to the output to minimize an objective function (Smola and Schölkopf, 2004; Chang and Lin, 2011). That is, the input/output mapping is assumed to nonlinear, but is constrained by the kernel being used (it is not completely flexible). In our toolbox, the default kernel is a radial basis function.
**XGBoost:** XGBoost (Extreme Gradient Boosting) (Chen and Guestrin, 2016) is an implementation of gradient boosted trees (Natekin and Knoll, 2013; Chen, 2014). Tree-based methods sequentially split the input space into many discrete parts (visualized as branches on a tree for each split), in order to assign each final "leaf" (a portion of input space that is not split any more) a value in output space (Breiman, 2017). XGBoost fits many regression trees, which are trees that predict continuous output values. "Gradient boosting" refers to fitting each subsequent regression tree to the residuals of the previous fit. This method assumes flexible nonlinear input/output mappings.

**Feedforward Neural Network:** A feedforward neural net (Rosenblatt, 1961; Goodfellow et al., 2016) connects the inputs to sequential layers of hidden units, which then connect to the output. Each layer connects to the next (e.g., the input layer to the first hidden layer, or the first to second hidden layers) via linear mappings followed by nonlinearities. Note that the Wiener cascade is a special case of a neural network with no hidden layers. Feedforward neural networks also allow flexible nonlinear input/output mappings.

**Simple RNN:** In a standard recurrent neural network (RNN) (Rumelhart et al., 1986; Goodfellow et al., 2016), the hidden state is a linear combination of the inputs and the previous hidden state. This hidden state is then run through an output nonlinearity, and linearly mapped to the output. Like feedforward neural networks, RNNs allow flexible nonlinear input/output mappings. Additionally, unlike feedforward neural networks, RNNs allow temporal changes in the system to be modeled explicitly.

**Gated Recurrent Unit:** Gated recurrent units (GRUs) (Cho et al., 2014; Goodfellow et al., 2016) are a more complex type of recurrent neural network. It has gated units, which determine how much information can flow through various parts of the network. In practice, these gated units allow for better learning of long-term dependencies.

**Long Short Term Memory Network:** Like the GRU, the long short term memory (LSTM) network (Hochreiter and Schmidhuber, 1997; Goodfellow et al., 2016) is a more complex recurrent neural network with gated units that further improve the capture of long-term dependencies. The LSTM has more parameters than the GRU.

**Kalman Filter:** Our Kalman filter for neural decoding was based on (Wu et al., 2003). In the Kalman filter, the hidden state at time $t$ is a linear function of the hidden state at time $t-1$, plus a matrix characterizing the uncertainty. For neural decoding, the hidden state is the kinematics (x and y components of position, velocity, and acceleration), which we aim to estimate. Note that even though we only aim to predict position or velocity, all kinematics are included because this allows for better prediction. Kalman filters assume that both the input/output mapping and the transitions in kinematics over time are linear.

**Naïve Bayes:** The Naïve Bayes decoder is a type of Bayesian decoder that determines the probabilities of different outcomes, and it then predicts the most probable. Briefly, it fits an encoding model to each neuron, makes conditional independence assumptions about neurons, and then uses Bayes' rule to create a decoding model from the encoding models. Thus, the effects of individual neurons are combined linearly. This probabilistic framework can incorporate prior information about the output variables, including the probabilities of their transitions over time. We used a Naïve Bayes decoder similar to the one implemented in (Zhang et al., 1998).

In the comparisons below, we also demonstrate an ensemble method. We combined the predictions from all decoders except the Kalman filter and Naïve Bayes decoders (which have different formats) using a feedforward neural network. That is, the eight methods' predictions were provided as input into a feedforward neural network that we trained to predict the true output.

### *Demonstration datasets*

We first examined which of several decoding methods performed the best across three datasets from motor cortex, somatosensory cortex, and hippocampus. All datasets are linked from our GitHub repository: https://github.com/KordingLab/Neural_Decoding .

.

In the task for decoding from motor cortex, monkeys moved a manipulandum that controlled a cursor on a screen (Glaser et al., 2018), and we aimed to decode the x and y velocity of the cursor. Details of the task can be found in (Glaser et al., 2018). The 21 minute recording from motor cortex contained 164 neurons. The mean and median firing rates, respectively, were 6.7 and 3.4 spikes / sec. Data were put into 50 ms bins. We used 700 ms of neural activity (the concurrent bin and 13 bins before) to predict the current movement velocity.

The same task was used in the recording from somatosensory cortex (Benjamin et al., 2018). The recording from S1 was 51 minutes and contained 52 neurons. The mean and median firing rates, respectively, were 9.3 and 6.3 spikes / sec. Data were put into 50 ms bins. We used 650 ms surrounding the movement (the concurrent bin, 6 bins before, and 6 bins after).

In the task for decoding from hippocampus, rats chased rewards on a platform (Mizuseki et al., 2009b; Mizuseki et al., 2009a), and we aimed to decode the rat's x and y position. From this recording, we used 46 neurons over a time period of 75 minutes. These neurons had mean and median firing rates of 1.7 and 0.2 spikes / sec, respectively. Data were put into 200 ms bins. We used 2 seconds of surrounding neural activity (the concurrent bin, 4 bins before, and 5 bins after) to predict the current position. Note that the bins used for decoding differed in all tasks for the Kalman filter and Naïve Bayes decoders (see *Extended Data* Figure 4-3 for additional details).

All datasets, which have been used in prior publications (Mizuseki et al., 2009b; Mizuseki et al., 2009a; Benjamin et al., 2018; Glaser et al., 2018), were collected with approval from the Institutional Animal Care and Use Committees of the appropriate institutions.

***Demonstration analysis methods***

**Scoring Metric:** To determine the goodness of fit, we used $R^2 = 1 - \frac{\sum_i (\hat{y}_i - y_i)^2}{\sum_i (y_i - \bar{y})^2}$, where $\hat{y}_i$ are the predicted values, $y_i$ are the true values and $\bar{y}$ is the mean value. This formulation of $R^2$ (Serruya et al., 2003; Fagg et al., 2009)(which is the fraction of variance accounted for, rather than the squared Pearson's correlation coefficient) can be negative on the test set due to overfitting on the training set. The reported $R^2$ values are the average across the x and y components of velocity or position.
**Preprocessing:** The training output was zero-centered (mean subtracted), except in support vector regression, where the output was normalized (z-scored), which helped algorithm performance. The training input was z-scored for all methods. The validation/testing inputs and outputs were preprocessed using the preprocessing parameters from the training set.
**Cross-validation:** When determining the $R^2$ for every method (Fig. 3), we used 10-fold cross-validation. For each fold, we split the data into a training set (80% of data), a contiguous validation set (10% of data), and a contiguous testing set (10% of data). For each fold, decoders were trained to minimize the mean squared error between the predicted and true velocities/positions of the training data. We found the algorithm hyperparameters that led to the highest $R^2$ on the validation set using Bayesian optimization (Snoek et al., 2012). That is, we fit many models on the training set with different hyperparameters and calculated the $R^2$ on the validation set. Then, using the hyperparameters that led to the highest validation set $R^2$, we calculated the $R^2$ value on the testing set. Error bars on the test set $R^2$ values were computed across cross-validation folds. Because the training sets on different folds were overlapping, computing the SEM as $\sigma/\sqrt{J}$ (where $\sigma$ is the standard deviation and $J$ is the number of folds) would have underestimated the size of the error bars (Nadeau and Bengio, 2000). We thus calculated the SEM as $\sigma * \sqrt{\frac{1}{J}+\frac{1}{J-1}}$ (Nadeau and Bengio, 2000), which takes into account that the estimates across folds are not independent.

**Bootstrapping:** When determining how performance scaled as a function of data size (Fig. 5), we used single test and validation sets, and varied the amounts of training data that directly preceded the validation set. We did not do this on 10 cross-validation folds due to long run-times. The test and validation sets were 5 minutes long for motor and somatosensory cortices, and 7.5 minutes for hippocampus. To get error bars, we resampled from the test set. Because of the high correlation between temporally adjacent samples, we didn't resample randomly from all examples (which would create highly correlated resamples). Instead, we separated the test set into 20 temporally distinct subsets, $S_1$-$S_{20}$ (i.e., $S_1$ is from $t=1$ to $t=T/20$, $S_2$ is from $t=T/20$ to $t=2T/20$, etc., where $T$ is the end time), to ensure that the subsets were more nearly independent of each other. We then resampled combinations of these 20 subsets (e.g., $S_5$, $S_{13}$, … $S_2$) 1000 times to get confidence intervals of $R^2$ values.

## Results

We investigated how the choice of machine learning technique affects decoding performance using a number of common machine learning methods that are included in our code package. These ranged from historical linear techniques (e.g., the Wiener filter) to modern machine learning techniques (e.g., neural networks and ensembles of techniques). We tested the performance of these techniques across datasets from motor cortex, somatosensory cortex, and hippocampus.

### *Performance comparison*

In order to get a qualitative impression of the performance, we first plotted the output of each decoding method for each of the three datasets (Fig. 3). In these examples, the modern methods, such as the LSTM and ensemble, appeared to outperform traditional methods. We next quantitatively compared the methods' performances, using the metric of $R^2$ on held-out test sets. These results confirmed our qualitative findings (Fig. 4). In particular, neural networks and the ensemble led to the best performance, while the Wiener or Kalman Filter led to the worst performance. In fact, the LSTM decoder explained over 40% of the unexplained variance from a Wiener filter ($R^2$'s of 0.88, 0.86, 0.62 vs. 0.78, 0.75, 0.35). Interestingly, while the Naïve Bayes decoder performed relatively well when predicting position in the hippocampus dataset (mean $R^2$ just slightly less than the neural networks), it performed very poorly when predicting hand velocities in the other two datasets. Another interesting finding is that the feedforward neural network did almost as well as the LSTM in all brain areas. Across cases, the ensemble method added a reliable, but small increase to the explained variance. Overall, modern ML methods led to significant increases in predictive power.

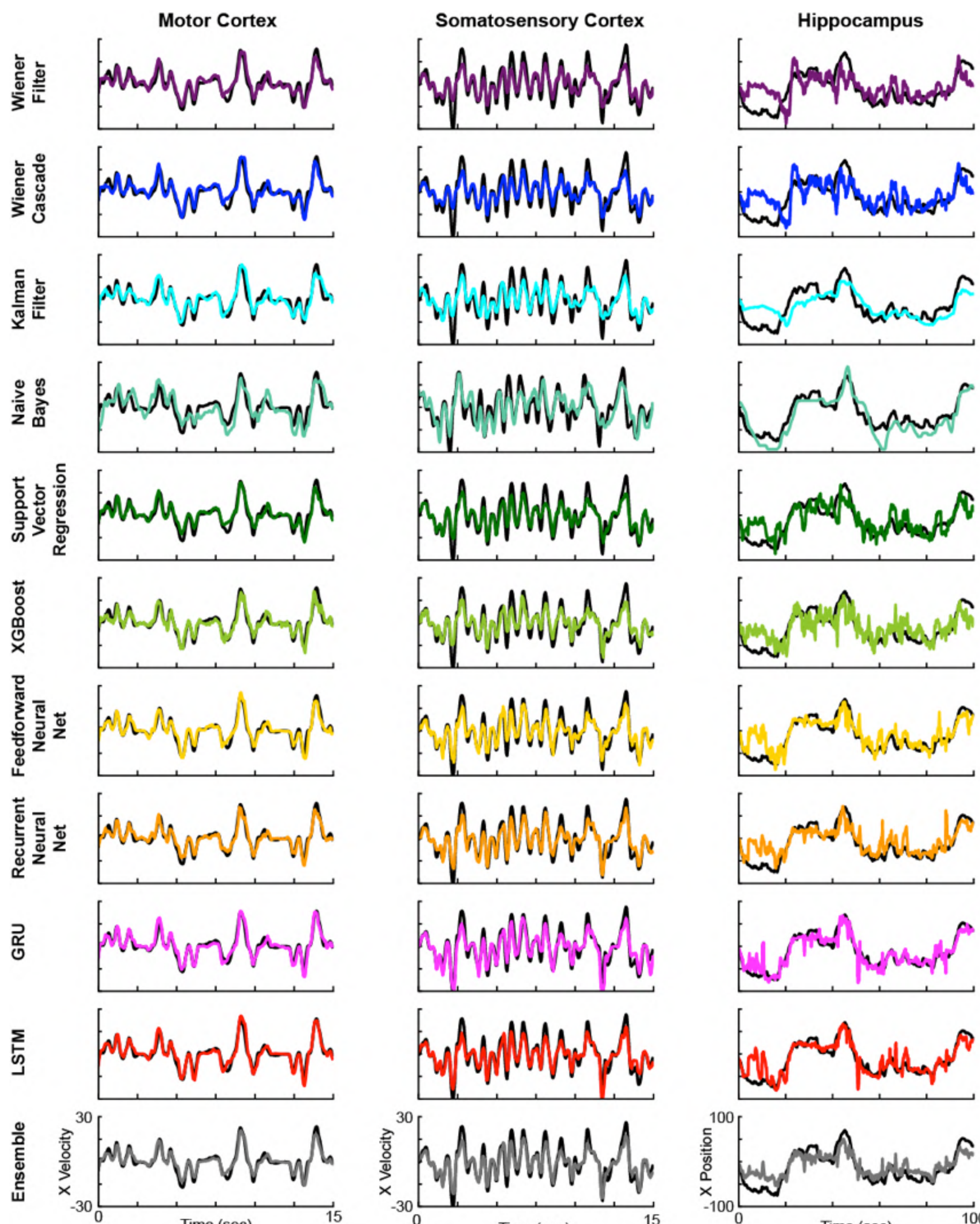

**Figure 3: Example Decoder Results**

Example decoding results from motor cortex (left), somatosensory cortex (middle), and hippocampus (right), for all eleven methods (top to bottom). Ground truth traces are in black, while decoder results are in various colors.

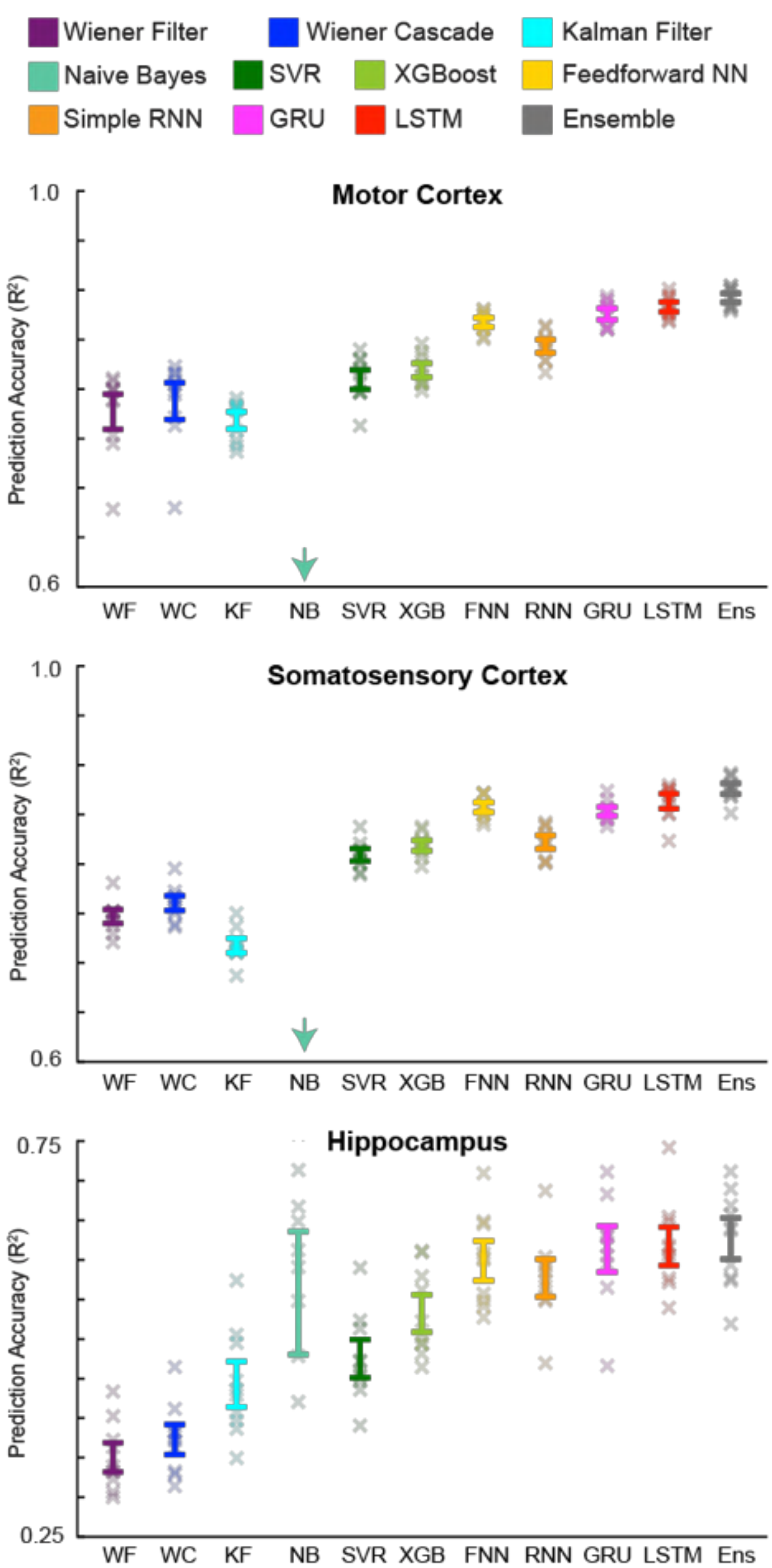

**Figure 4: Decoder Result Summary**

$R^2$ values are reported for all decoders (different colors) for each brain area (top to bottom). Error bars represent the mean +/- SEM across cross-validation folds. X's represent the $R^2$ values of each cross-validation fold. The NB decoder had mean $R^2$ values of 0.26 and 0.36 (below the minimum y-axis value) for the motor and somatosensory cortex datasets, respectively. Note the different y-axis limits for the hippocampus dataset in this and all subsequent figures. In Extended Data, we include the accuracy for multiple versions of the Kalman filter (Figure 4-1), accuracy for multiple bin sizes (Figure 4-2), and a table with further details of all these methods (Figure 4-3).

***Concerns about limited data for decoding***

We chose a representative subset of the ten methods to pursue further questions about particular aspects of neural data analysis: the feedforward neural network and LSTM (two effective modern methods), along with the Wiener and Kalman filters (two traditional methods in widespread use). The improved predictive performance of the modern methods is likely due to their greater complexity. However, this greater complexity may make these methods unsuitable for smaller amounts of data. Thus, we tested performance with varying amounts of training data. With only 2 minutes of data for motor and somatosensory cortices, and 15 minutes of hippocampus data, both modern methods outperformed both traditional methods (Fig. 5a, Extended Data Fig. 5-1). When decreasing the amount of training data further, to only 1 minute for motor and somatosensory cortices and 7.5 minutes for hippocampus data, the Kalman filter performance was sometimes comparable to the modern methods, but the modern methods significantly outperformed the Wiener Filter (Fig. 5a). Thus, even for limited recording times, modern ML methods can yield significant gains in decoding performance.

Besides limited recording times, neural data is often limited in the number of recorded neurons. Thus, we compared methods using a subset of only 10 neurons. For motor and somatosensory data, despite a general decrease in performance for all decoding methods, the modern methods significantly outperformed the traditional methods (Fig. 5b). For the hippocampus dataset, no method predicted well (mean $R^2 < 0.25$) with only 10 neurons. This is likely because 10 sparsely firing neurons (median firing of HC neurons was ~0.2 spikes / sec) did not contain enough information about the entire space of positions. However, in most scenarios, with limited neurons and for limited recorded times, modern ML methods can be advantageous.

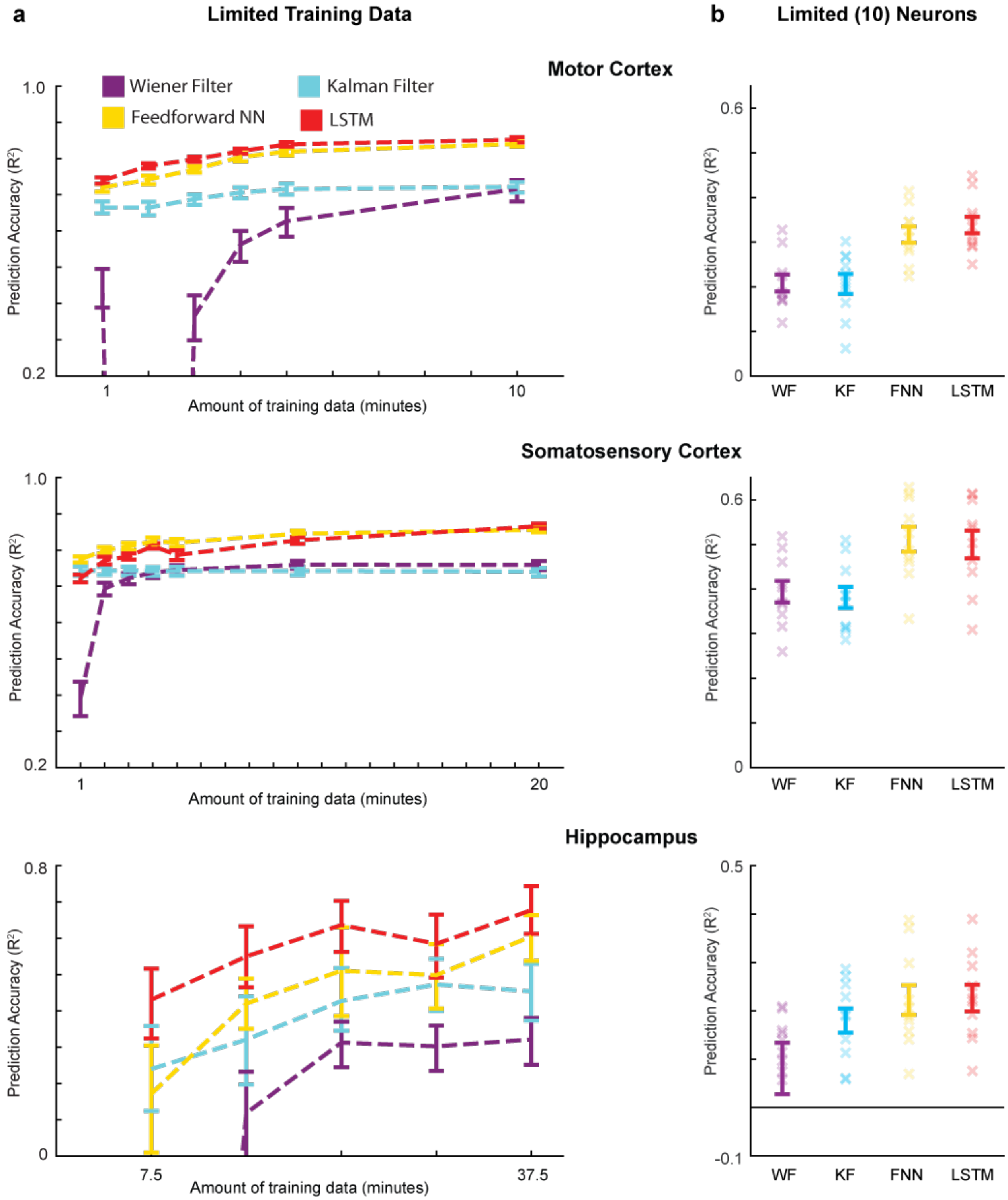

**Figure 5: Decoder results with limited data**
**(A)** Testing the effects of limited training data. Using varying amounts of training data, we trained two traditional methods (Wiener filter and Kalman filter), and two modern methods (feedforward neural network and LSTM). $R^2$ values are reported for these decoders (different colors) for each brain area (top to bottom). Error bars are 68% confidence intervals (meant to approximate the SEM) produced via bootstrapping, as we used a single test set. Values with negative $R^2$s were not shown. **(B)** Testing the effects of few neurons. Using only 10 neurons, we trained two traditional methods (Wiener filter and Kalman filter), and two modern methods (feedforward neural network and LSTM). We used the same testing set as in panel A, and the largest training set from panel A. $R^2$ values are reported for these decoders for each brain area. Error bars represent the mean +/- SEM of multiple repetitions with different subsets of 10 neurons. X's represent the $R^2$ values of each repetition. Note that the y-axis limits are different in panels A and B. In Extended Data, we provide examples of the decoder predictions for each of these methods (Figure 5-1).

### *Concerns about run-time*

While modern ML methods lead to improved performance, it is important to know that these sophisticated decoding models can be trained in a reasonable amount of time. To get a feeling for the typical timescale, consider that when running our demonstration data on a desktop computer using only CPUs, it took less than 1 second to fit a Wiener filter, less than 10 seconds to fit a feedforward neural, and less than 8 minutes to fit an LSTM. (This was for 30 minutes of data, using 10 time bins of neural activity for each prediction, and 50 neurons.) In practice, these models will need to be fit tens to hundreds of times when incorporating hyperparameter optimization and cross-validation. As a concrete example, let us say that we are doing 5-fold cross-validation, and hyperparameter optimization requires 50 iterations per cross-validation fold. In that case, fitting the LSTM would take about 30 hours, and fitting the feedforward neural net would take less than 1 hour. Note that variations in hardware and software implementations can drastically change runtime, and modern GPUs can often increase speeds approximately 10-fold. Thus, while modern methods do take significantly longer to train (and may require running overnight without GPUs), that should be manageable for most offline applications.

### *Concerns about robustness to hyperparameters*

All our previous results used hyperparameter optimization. While we strongly encourage a thorough hyperparameter optimization, a user with limited time might just do a limited hyperparameter search. Thus, it is helpful to know how sensitive results may be to varying hyperparameters. We tested the performance of the feedforward neural network while varying two hyperparameters: the number of units and the dropout rate (a regularization hyperparameter for neural networks). We held the third hyperparameter in our code package, the number of training epochs, constant at 10. We found that the performance of the neural network was generally robust to large changes in the hyperparameters (Fig. 6). As an example, for the somatosensory cortex dataset, the peak performance of the neural network was $R^2$=0.86 with 1000 units and 0 dropout, and virtually the same ($R^2$=0.84) with 300 units and 30% dropout. Even when using limited data, neural network performance was robust to hyperparameter changes. For instance, when training the somatosensory cortex dataset with 1 minute of training data, the peak performance was $R^2$=0.77 with 700 units and 20% dropout. A network with 300 units and 30% dropout had $R^2$=0.75. Note that the hippocampus dataset, in particular when using limited training data, did have greater variability, emphasizing the importance of hyperparameter optimization on sparse datasets. However, for most datasets, researchers should not be concerned that slightly non-optimal hyperparameters will lead to largely degraded performance.

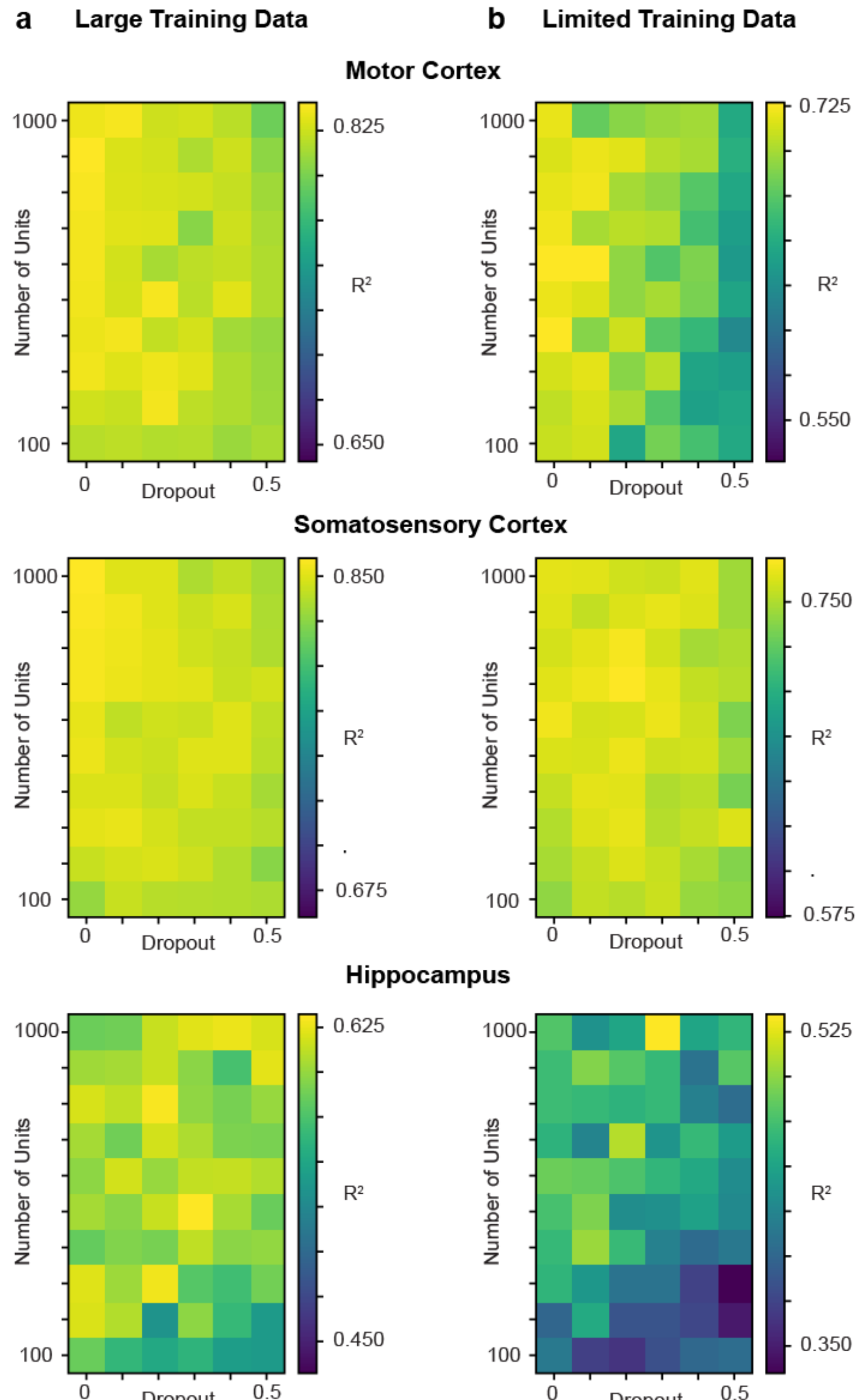

**Figure 6: Sensitivity of neural network results to hyperparameter selection**

In a feedforward neural network, we varied the number of hidden units per layer (in increments of 100) and the proportion of dropout (in increments of 0.1), and evaluated the decoder's performance on all three datasets (top to bottom). The neural network had two hidden layers, each with the same number of hidden units. The number of training epochs was kept constant at 10. The colors show the $R^2$ on the test set, and each panel's colors were put in the range: [maximum $R^2$ – 0.2 , maximum $R^2$]. **a**) We used a large amount of training data (the maximum amount used in Fig. 5a), which was 10, 20, and 37.5 minutes of data for the motor cortex, somatosensory cortex, and hippocampus datasets, respectively. **b**) Same results for a limited amount of training data: 1, 1, and 15 minutes of data for the motor cortex, somatosensory cortex, and hippocampus datasets, respectively.

## Discussion

Here we have provided a tutorial, code package, and demonstrations of the use of machine learning for neural decoding. Our comparisons, which were made using the code we have published online, show that machine learning works well on typical neural decoding datasets, outperforming traditional decoding methods. In our demonstration, we decoded continuous-valued variables. However, these same methods can be used for classification tasks, which often use classic decoders such as logistic regression and support vector machines. Our available code also includes classification methods.

We find it particularly interesting that the neural network methods worked so well with limited data, counter to the common perception. We believe the explanation is simply the size of the networks. For instance, our networks have on the order of $10^5$ parameters, while common networks for image classification (e.g., (Krizhevsky et al., 2012)) can have on the order of $10^8$ parameters. Thus, the smaller size of our networks (hundreds of hidden units) may have allowed for excellent prediction with limited data (Zhang et al., 2016). Moreover, the fact that the tasks we used had a low-dimensional structure, and therefore the neural data was also likely low dimensional (Gao et al., 2017), might allow high decoding performance with limited data.

In order to find the best hyperparameters for the decoding algorithms, we used a Bayesian optimization routine (Snoek et al., 2012) to search the hyperparameter space (see *Demonstration Methods*). Still, it is possible that, for some of the decoding algorithms, the hyperparameters were nonoptimal, which would have lowered overall accuracy. Moreover, for several methods, we did not attempt to optimize all the hyperparameters. We did this in order to simplify the use of the methods, decrease computational runtime during hyperparameter optimization, and because optimizing additional hyperparameters (beyond default values) did not appear to improve accuracy. For example, for the neural nets we used dropout but not L1 or L2 regularization, and for XGBoost we optimized less than half the available hyperparameters designed to avoid overfitting. While our preliminary testing with additional hyperparameters did not appear to change the results significantly, it is possible that some methods did not achieve optimal performance.

We have decoded from spiking data, but it is possible that the problem of decoding from other data modalities is different. One main driver of a difference may be the distinct levels of noise. For example, fMRI signals have far higher noise levels than spikes. As the noise level goes up, linear techniques become more appropriate, which may ultimately lead to a situation where the traditional linear techniques become superior. Applying the same analyses we did here across different data modalities is an important next step.

All our decoding was done "offline," meaning that the decoding occurred after the recording, and was not part of a control loop. This type of decoding is useful for determining how information in a brain area relates to an external variable. However, for engineering applications such as BMIs (Nicolas-Alonso and Gomez-Gil, 2012; Kao et al., 2014), the goal is to decode information (e.g., predict movements) in real time. Our results here may not apply as directly to online decoding situations, since the subject is ultimately able to adapt to imperfections in the decoder. In that case, even relatively large decoder performance differences may be irrelevant. Plus, there are additional challenges in online applications, such as non-stationary inputs (e.g., due to electrodes shifting in the brain) (Wu and Hatsopoulos, 2008; Sussillo et al., 2016; Farshchian et al., 2018). Finally, online applications are concerned with computational runtime, which we have only briefly addressed here. In the future, it would be valuable to test modern techniques for decoding in online applications (as in (Sussillo et al., 2012; Sussillo et al., 2016)).

Finally, we want to briefly mention the concept of feature importance, which our tutorial has not addressed. Feature importance refers to the determination of which inputs most affect a machine learning model's predictions. For decoding, feature importance methods could be used to ask which neurons are important for making the predictions, or if multiple brain areas are input into the decoder, which of these brain regions matter. Three common, straightforward approaches that we would recommend are 1) to build separate decoders with individual features to test those features' predictive abilities; 2) leave-one-feature-out, in which the decoder is fit while leaving out a feature, in order to evaluate how removing that feature decreases prediction performance; and 3) permutation feature importance, in which after fitting the decoder, a feature's values are shuffled to determine how much prediction performance decreases. (Molnar, 2018; Dankers et al., 2020) provide more details on these approaches, and (Molnar, 2018) describes other common feature importance methods in the machine learning literature. These methods can facilitate some understanding of neural activity even when the ML decoder is quite complex.

**Conclusion**

Machine learning is relatively straightforward to apply and can greatly improve the performance of neural decoding. Its principle advantage is that many fewer assumptions need to be made about the structure of the neural activity and the decoded variables. Best practices like testing on held-out data, perhaps via crossvalidation, are crucial to the ML pipeline and are critical for any application. The Python package that accompanies this tutorial is designed to guide both best practices and the deployment of specific ML algorithms, and we expect it will be useful in improving decoding performance for new datasets. Our hunch is that it will be hard for specialized algorithms (e.g. (Corbett et al., 2010; Kao et al., 2017)) to compete with the standard algorithms developed by the machine learning community.

## Acknowledgements

We would like to thank Pavan Ramkumar for help with code development. For funding, JG was supported by NIH F31 EY025532 and NIH T32 HD057845, NSF NeuroNex Award DBI-1707398, and the Gatsby Charitable Foundation. AB was supported by NIH MH103910. MP was supported by NIH F31 NS092356 and NIH T32 HD07418. RC was supported by NIH R01 NS095251 and DGE-1324585. LM was supported by NIH R01 NS074044 and NIH R01 NS095251. KK was supported by NIH R01 NS074044, NIH R01 NS063399 and NIH R01 EY021579.

# Extended Data

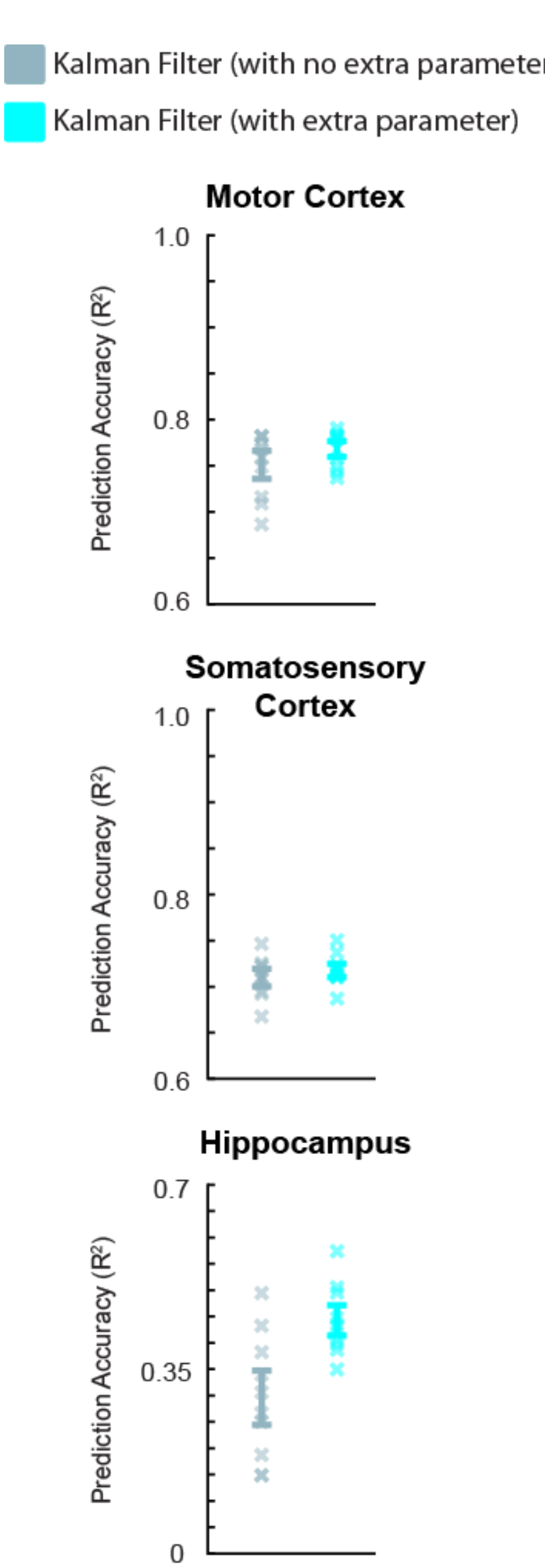

**Figure 4-1. Kalman Filter Versions**

$R^2$ values are reported for different versions of the Kalman Filter for each brain area (top to bottom). On the left (in bluish gray), the Kalman Filter is implemented as in (Wu et al., 2003). On the right (in cyan), the Kalman Filter is implemented with an extra parameter that scales the noise matrix associated with the transition in kinematic states (see *Demonstration Methods*). This version with the extra parameter is the one used in the main text. Error bars represent the mean +/- SEM across cross-validation folds. X's represent the $R^2$ values of each cross-validation fold. Note the different y-axis limits for the hippocampus dataset.

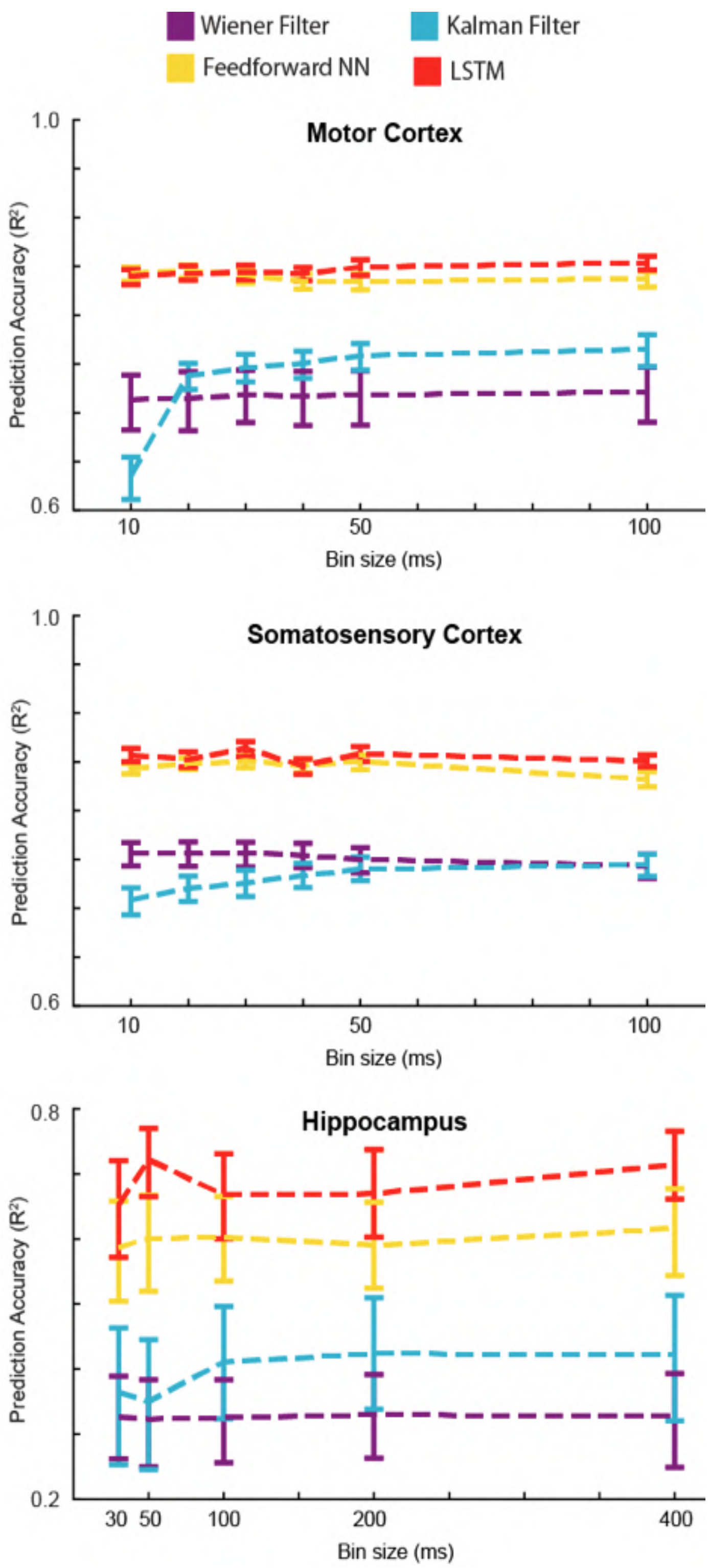

**Figure 4-2: Decoder results with different bin sizes**

As different decoding applications may require different temporal resolutions, we tested a subset of methods with varying bin sizes. We trained two traditional methods (Wiener filter and Kalman filter), and two modern methods (feedforward neural network and LSTM). We used the same testing set as in Fig. 5, and the largest training set from Fig. 5. $R^2$ values are reported for these decoders (different colors) for each brain area (top to bottom). Error bars are 68% confidence intervals (meant to approximate the SEM) produced via bootstrapping, as we used a single test set. Modern machine learning methods remained advantageous regardless of the temporal resolution.

Note that for this figure, we used a slightly different amount of neural data than in other analyses, in order to have a quantity that was divisible by many bin sizes. In this case, for motor cortex, we used 600 ms of neural activity prior to and including the current bin. For somatosensory cortex, we used 600 ms of neural activity centered on the current bin. For hippocampus, we used 2 seconds of neural activity, centered on the current bin.

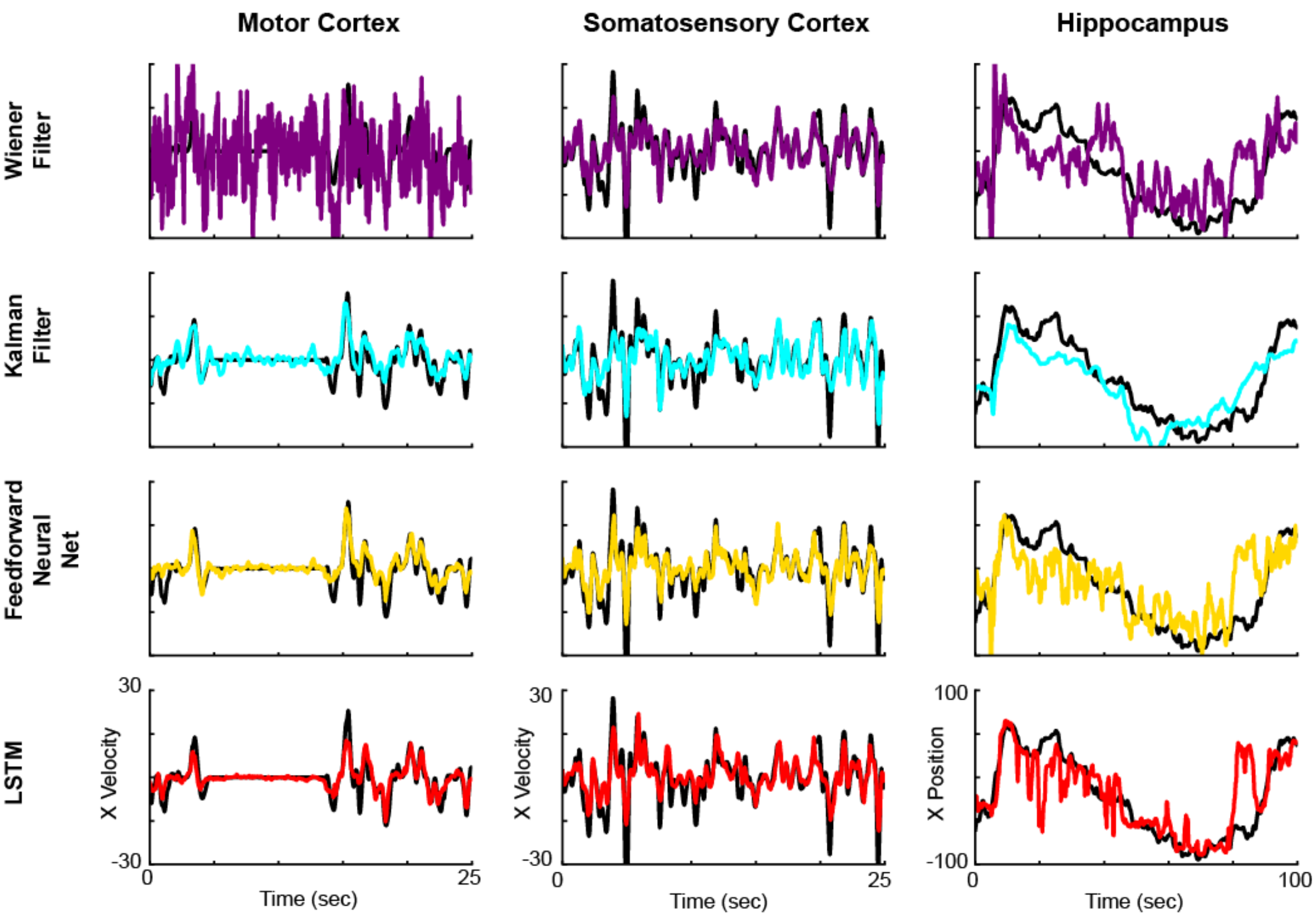

**Figure 5-1: Example results with limited training data**

Using only 2 minutes of training data for motor cortex and somatosensory cortex, and 15 minutes of training data for hippocampus, we trained two traditional methods (Wiener filter and Kalman filter), and two modern methods (feedforward neural network and LSTM). Example decoding results are shown from motor cortex (left), somatosensory cortex (middle), and hippocampus (right), for these methods (top to bottom). Ground truth traces are in black, while decoder results are in the same colors as previous figures.

| Method | Equations, details, and hyperparameters |
|---|---|
| *General information* | For all the methods below, we describe *our particular uses*, which are available in our code package and are used in the decoding demonstrations we show in *Results*.<br><br>Let $\boldsymbol{X}$ be the input matrix of covariates and let $\boldsymbol{Y}$ be the output that is being predicted. $\boldsymbol{Y}$ is either a vector or matrix depending on whether the outputs (e.g. x and y components of velocity) are predicted simultaneously.<br><br>Also, note that for our demonstrations, for all methods below, the noise of the output is assumed to be normally distributed. |
| Wiener Filter | Decoder predictions have the form $\boldsymbol{Y} = \boldsymbol{\theta}\boldsymbol{X}$, where $\boldsymbol{\theta}$ is fit according to maximum likelihood estimation.<br>*Hyperparameters:* None |
| Wiener Cascade | Decoder predictions have the form $\boldsymbol{Y} = \beta_0 + \beta_1\boldsymbol{Z}^1 + \cdots + \beta_n\boldsymbol{Z}^n$, where $\boldsymbol{Z} = \boldsymbol{\theta}\boldsymbol{X}$. The parameters $\boldsymbol{\theta}$ and the $\beta$'s are fit iteratively according to maximum likelihood estimation.<br>*Hyperparameters:* Degree of the polynomial used for the nonlinearity, *n.* |
| SVR | Tutorials have previously been written on the details of support vector regression and how they are optimized (Smola and Schölkopf, 2004). Here, we provide a very short mathematical summary. In nonlinear SVR:<br><br>$$\boldsymbol{Y} = \sum_{i=1}^{l}(\alpha_i - \alpha_i^*)k(\boldsymbol{X_i}, \boldsymbol{X}) + b$$<br><br>where *i* represents the data point. *k* is the kernel function, which allows for the nonlinear transformation. We used a radial basis function kernel:<br><br>$$k(\boldsymbol{x}', \boldsymbol{x}) = \exp\left(-\frac{\lVert\boldsymbol{x}' - \boldsymbol{x}\rVert^2}{\boldsymbol{\sigma}^2}\right)$$<br><br>The parameters $\alpha_i$ and $\alpha_i^*$ are constrained to be in the range $[0, C]$ in the optimization problem, where $C$ is a hyperparameter that can be understood as being inversely related to regularization strength.<br>*Hyperparameters:* Penalty of the error term ($C$), maximum number of iterations |
| XGBoost | In gradient-boosted tree methods (which XGBoost is an implementation of), a mapping is learned using a collection of regression trees. Training an individual regression tree is accomplished by sequentially splitting the input space, and assigning an output value to each final partition of the input space (see (Breiman, 2017) for further description). Let $f_i$ be the input/output mapping of regression tree *i.* The final decoder, which combines across *N* regression trees, will have the form:<br><br>$$\boldsymbol{Y} = \sum_{k=1}^{N} f_k(\boldsymbol{X})$$<br><br>To train the XGBoost decoder, regression trees are trained sequentially. The first tree is trained to minimize the loss function which here is $\lVert\boldsymbol{Y} - f_1\boldsymbol{X}\rVert_2^2 + \Omega(f_1)$, where $\Omega$ is a regularization function. The second tree is then trained to predict the residual of the first tree's predictions. That is, we optimize $f_2$ to minimize $\lVert\boldsymbol{Y} - f_1\boldsymbol{X} - f_2\boldsymbol{X}\rVert_2^2 + \Omega(f_2)$. Successive trees continue to be optimized to predict the remaining residual. See https://xgboost.readthedocs.io/en/latest/tutorials/model.html for a more in-depth tutorial description. |
| | *Hyperparameters:* Maximum depth of each regression tree, number of trees, learning rate |
| Feedforward Neural Net | We created a fully connected (dense) feedforward neural network with 2 hidden layers and rectified linear unit ("ReLU") (Glorot et al., 2011) activations after each hidden layer. We required the number of hidden units in each layer to be the same. This can be written as:<br><br>$\boldsymbol{H_1} = \mathrm{ReLU}(\boldsymbol{W_1 X} + \boldsymbol{b_1})$<br>$\boldsymbol{H_2} = \mathrm{ReLU}(\boldsymbol{W_2 X} + \boldsymbol{b_2})$<br>$\boldsymbol{Y} = \boldsymbol{W_3 H_2} + \boldsymbol{b_3}$<br><br>We used the Adam algorithm (Kingma and Ba, 2014) as the optimization routine to solve for all $\boldsymbol{W}$'s and $\boldsymbol{b}$'s. We used the Keras library (Chollet, 2015) for all neural network implementations. This neural network, and all neural networks below had two output units. That is, the same network predicted the x and y components (of position or velocity) together, rather than separately.<br>*Hyperparameters:* Number of units (dimensionality of $\boldsymbol{H_1}$ and $\boldsymbol{H_2}$), amount of dropout during training, number of training epochs |
| Simple RNN | Decoder predictions have the form*:*<br><br>$\boldsymbol{H_t} = \mathrm{ReLU}(\boldsymbol{U X_t} + \boldsymbol{W H_{t-1}} + \boldsymbol{b})$<br>$\boldsymbol{Y_t} = \boldsymbol{V H_t} + \boldsymbol{c}$<br><br>Note that we used the ReLU nonlinearity as opposed to the "tanh" nonlinearity, which is more common in RNNs, as this improved performance. We used RMSprop (Tieleman and Hinton, 2012) as the optimization routine.<br>*Hyperparameters:* Number of units (dimensionality of $\boldsymbol{H}$), amount of dropout during training, number of training epochs |
| GRUs | Here, we mix notation from (Goodfellow et al., 2016) and (Olah, 2015), which provide excellent descriptions of the method. The GRU has two gates, which control the transmission of information through the network. The "update" and "reset" gates are parameterized as follows:<br><br>Update gate: $\boldsymbol{z_t} = \sigma(\boldsymbol{U_z X_t} + \boldsymbol{W_z H_{t-1}} + \boldsymbol{b_z})$<br>Reset gate: $\boldsymbol{r_t} = \sigma(\boldsymbol{U_r X_t} + \boldsymbol{W_r H_{t-1}} + \boldsymbol{b_r})$<br><br>A temporary hidden state is calculated as:<br><br>$\boldsymbol{\widetilde{h}_t} = \tanh(\boldsymbol{U_h X_t} + \boldsymbol{r_t} * \boldsymbol{W_h H_{t-1}} + \boldsymbol{b_c})$<br><br>In this calculation, the reset gate controls what information from the previous hidden state is in this proposed new hidden state. Note that the $*$ symbol denotes an elementwise (by neural network unit) multiplication. To calculate the hidden state, we use the update gate to control what information flows from the previous hidden state versus what is updated from the temporary state:<br><br>$\boldsymbol{h_t} = (\boldsymbol{1} - \boldsymbol{z_t}) * \boldsymbol{h_{t-1}} + \boldsymbol{z_t} * \boldsymbol{\widetilde{h}_t}$<br><br>Finally, as in the standard RNN:<br>$\boldsymbol{Y_t} = \boldsymbol{V H_t} + \boldsymbol{c}$<br><br>We used RMSprop (Tieleman and Hinton, 2012) as the optimization routine. |
| | *Hyperparameters:* Number of units (dimensionality of $\boldsymbol{H}$), amount of dropout during training, number of training epochs. |
| LSTM | As above, we mix notation from (Goodfellow et al., 2016) and (Olah, 2015), which provide excellent descriptions of the method. The LSTM has three gates, which control the transmission of information through the network. The “forget”, “input”, and “output” gates are parameterized as follows:<br><br>$\boldsymbol{f_t} = \sigma(\boldsymbol{U_f X_t} + \boldsymbol{W_f H_{t-1}} + \boldsymbol{b_f})$<br>$\boldsymbol{i_t} = \sigma(\boldsymbol{U_i X_t} + \boldsymbol{W_i H_{t-1}} + \boldsymbol{b_i})$<br>$\boldsymbol{o_t} = \sigma(\boldsymbol{U_o X_t} + \boldsymbol{W_o H_{t-1}} + \boldsymbol{b_o})$<br><br>The LSTM has a cell state, $\boldsymbol{C_t}$, that carries information across time. We can calculate $\boldsymbol{C_t}$, as:<br><br>$\widetilde{\boldsymbol{C}}_{\boldsymbol{t}} = \tanh(\boldsymbol{U_c X_t} + \boldsymbol{W_c H_{t-1}} + \boldsymbol{b_c})$<br>$\boldsymbol{C_t} = \boldsymbol{f_t} * \boldsymbol{C_{t-1}} + \boldsymbol{i_t} * \widetilde{\boldsymbol{C}}_{\boldsymbol{t}}$<br><br>where $\widetilde{\boldsymbol{C}}_{\boldsymbol{t}}$ is a temporary cell state, that becomes part of the cell state depending on the input gate, $\boldsymbol{i_t}$. Additionally, the previous time point’s cell state is carried through depending on the forget gate, $\boldsymbol{f_t}$. Note that the $*$ symbol denotes an elementwise (by neural network unit) multiplication.<br><br>The hidden state of the network is an output-gated version of the cell state:<br>$\boldsymbol{H_t} = \boldsymbol{o_t} * \tanh(\boldsymbol{C_t})$<br><br>Finally, as in the standard RNN:<br>$\boldsymbol{Y_t} = \boldsymbol{V H_t} + \boldsymbol{c}$<br><br>We used RMSprop (Tieleman and Hinton, 2012) as the optimization routine.<br><br>*Hyperparameters:* Number of units (dimensionality of $\boldsymbol{H}$), amount of dropout during training, number of training epochs |
| Kalman Filter | In the standard Kalman Filter,<br><br>$\boldsymbol{y}_t = \boldsymbol{A}\boldsymbol{y}_{t-1} + \boldsymbol{w}$<br><br>where $\boldsymbol{y}_t$ and $\boldsymbol{y}_{t-1}$ are 6 x 1 vectors reflecting kinematic variables (not just the velocity or position being predicting), and $\boldsymbol{w}$ is sampled from a normal distribution, $N(0,\boldsymbol{W})$, with mean 0 and covariance $\boldsymbol{W}$.<br><br>$\boldsymbol{x}_{t*} = \boldsymbol{H}\boldsymbol{y}_t + \boldsymbol{q}$<br><br>where $\boldsymbol{x}_{t*}$ is the neural activity at time $t^*$, and $\boldsymbol{q}$ is sampled from a normal distribution, $N(0,\boldsymbol{Q})$. Note that we allowed a lag between the neural data and predicted kinematics, which is why we use $t^*$ for the time of the neural activity.<br><br>During training, $\boldsymbol{A}$, $\boldsymbol{H}$, $\boldsymbol{W}$, and $\boldsymbol{Q}$ are empirically fit on the training set using maximum likelihood estimation. When making predictions, to update the estimated hidden state at a given time point, the updates derived from the current measurement and the previous hidden states are combined. During this combination, the noise matrices give a higher weight to the less uncertain information. See (Wu et al., 2003) or our code for the update equations (note that $\boldsymbol{x}$ and $\boldsymbol{y}$ have different notation in (Wu et al., 2003)).<br><br>We had one hyperparameter which differed from the standard implementation (Wu et al., |
| | 2003). We divided the noise matrix associated with the transition in kinematic states, ***W,*** by the hyperparameter scalar *C*, which allowed weighting the neural evidence and kinematic transitions differently. The rationale for this addition is that neurons have temporal correlations, which make it desirable to have a parameter that allows changing the weight of the new neural evidence. The introduction of this parameter made a big difference for the hippocampus dataset (Extended Data Fig. 4 - 1).<br><br>*Hyperparameters: C* and the lag between the neural data and predicted kinematics. |
| Naïve Bayes | We used a Naïve Bayes decoder similar to the one implemented in (Zhang et al., 1998). We first fit an encoding model (tuning curve) using the output variables. Let $f_i(\mathbf{s})$ be the value of the tuning curve (the expected number of spikes) for neuron *i* at the output variables ***s***. Note that ***s*** is a vector containing the two output variables we are predicting (x and y positions/velocities). We assume the number of recorded spikes in the given bin, $r_i$, is generated from the tuning curve with Poisson statistics:<br><br>$$P(r_i \mid \mathbf{s}) = \frac{\exp\,[-f_i(\mathbf{s})]\, f_i(\mathbf{s})^{r_i}}{r_i!}$$<br><br>We also assume that all the neurons' spike counts are conditionally independent given the output variables, so that:<br><br>$$P(\mathbf{r} \mid s) \propto \prod_i P(r_i \mid s)$$<br><br>where ***r*** is a vector with the spike counts of all neurons. Bayes' rule can then be used to determine the likelihood of the output variables given the spike counts of all neurons:<br><br>$$P(\mathbf{s} \mid \mathbf{r}) \propto\ P(\mathbf{r} \mid \mathbf{s})P(\mathbf{s})$$<br><br>where $P(\mathbf{s})$ is the probability distribution of the output variables. To help with temporal continuity of decoding, we want our probabilistic model to include how the output variables at one time step depend on the output variables at the previous time step: $P(\mathbf{s}_t \mid \mathbf{s}_{t-1})$. Thus, we can more generally write, using Bayes' rule as before:<br><br>$$P(\mathbf{s}_t \mid \mathbf{r}_{t*}, \mathbf{s}_{t-1}) \propto\ P(\mathbf{r}_{t*} \mid \mathbf{s}_t)P(\mathbf{s}_{t-1} \mid \mathbf{s}_t)P(\mathbf{s}_t)$$<br><br>Note that we use $\mathbf{r}_{t*}$ rather than $\mathbf{r}_t$ because we use neural responses from multiple time bins to predict the current output variables. The above formula assumes that $\mathbf{r}_{t*}$ and $\mathbf{s}_{t-1}$ are independent, conditioned on $\mathbf{s}_t$. The final decoded stimulus in a time bin is: $\mathrm{argmax}_{\mathbf{s}_t}\ P(\mathbf{s}_t \mid \mathbf{r}_{t*}, \mathbf{s}_{t-1})$.<br><br>$P(\mathbf{s}_{t-1} \mid \mathbf{s}_t)$ was determined as follows. Let $\Delta\mathbf{s}$ be the Euclidean distance in ***s*** from one time step to the next. We fit $P(\Delta\mathbf{s})$ as a Gaussian using data from the training set. $P(\mathbf{s}_{t-1} \mid \mathbf{s}_t)$ was approximated as $P(\Delta\mathbf{s}_t)$. That is, the probability of going from one output state to another was only based on the distance between the output states, not the output state itself.<br><br>Additionally, including $P(\mathbf{s})$ based on the distribution of output variables in the training set did not improve performance on the validation set. This could be because the probability distribution differed between the training and validation/testing sets, or because the distribution of output variables was approximately uniform in our tasks. Thus, we simply used a uniform prior. |
| | In our calculations, we discretize $\boldsymbol{s}$ into a 100 x 100 grid going from the minimum to maximum of the output variables. When increasing the decoding resolution of the output variables, we did not see a meaningful change in decoding accuracy.<br><br>Our tuning curves had the format of a Poisson generalized quadratic model (Park et al., 2013), which improved the performance over generalized linear models on validation datasets.<br><br>On the hippocampus dataset, we used the total number of spikes over the same time interval as we used for the other decoders (4 bins before, the concurrent bin, and 5 bins after). Note that using a single time bin of spikes led to very poor performance. On the motor cortex and somatosensory cortex datasets, the naïve Bayes decoder gave very poor performance regardless of the bins used. We ultimately used bins that gave the best performance on a validation set: 2 bins before and the concurrent bin for the motor cortex dataset; 1 bin before, the concurrent bin, and 1 bin after for the somatosensory cortex dataset.<br><br>*Hyperparameters:* None |

**Figure 4-3: Additional decoder details, including equations and hyperparameters**
These details are for the decoder implementations that we use in our demonstrations and have in our code package.